\documentstyle[12pt,epsfig]{article}
\def\lapp{{\ \lower 0.6ex \hbox{$\buildrel<\over\sim$}\ }}
\def\gapp{{\ \lower 0.6ex \hbox{$\buildrel>\over\sim$}\ }}
\begin{document}
\begin{titlepage}
\vspace*{-1cm}
\begin{flushright}
FTUV/98/47    \\
IFIC/98/48   \\
June 1998 \\
\end{flushright}                                
\vskip 1.cm
\begin{center}                                                                  
{\Large\bf
Three--body decays of Higgs bosons at LEP2 and\\
application to a hidden fermiophobic Higgs.}
 \vskip 1.cm
{\large A.G.~Akeroyd\footnote{akeroyd@flamenco.ific.uv.es \hspace*{0.5cm} 
Tel/Fax 34-96-3864555}}
\vskip .4cm
 Departamento de F\'\i sica Te\'orica, IFIC/CSIC,\\
 Universidad de Valencia, Burjassot 46100,\\
Valencia, Spain
\vskip 1cm                                                                    
\end{center}                                                                    

\begin{abstract}

We study the decays of Higgs bosons to a lighter
Higgs boson and a virtual gauge boson in the context of the
non--supersymmetric Two--Higgs--Doublet--Model (2HDM). We consider the
phenomenological impact at LEP2 and find that such decays, 
when open, may be dominant in regions of parameter space and thus
affect current Higgs boson search techniques. 
Three--body decays would be a way of producing light neutral Higgs bosons
which have so far escaped detection at LEP due to suppressed couplings
to the $Z$, and are of particular importance in the 2HDM (Model~I)
 which allows both a light fermiophobic Higgs and a light charged scalar.

\end{abstract}
\hspace*{1.0cm}

\vfill
\end{titlepage}                                                                 
\newpage                                                                        

\section{Introduction}
The Higgs sector \cite{Higgs} of the Standard Model (SM)
\cite{Wein} is still experimentally
untested, and so far only a lower bound on the mass of the Higgs boson 
($M_{\phi^{\circ}}\ge 87.6$ GeV)
has been obtained \cite{moriond}. The minimal SM possesses one complex scalar
doublet with a non--zero vacuum expectation value (VEV), and after
symmetry breaking predicts a neutral Higgs boson.  
Enlarged Higgs sectors with $N$ doublets may be considered \cite{Gun},
\cite{Gross} and 
predict charged Higgs bosons ($H^{\pm}$) and additional
neutral scalars. 
Accurate predictions of the branching ratios of these particles are needed
in order to facilitate the searches at future colliders, and the present 
work considers
decays of a Higgs boson to a lighter Higgs boson and a virtual vector boson.
 We shall be focusing on the Two--Higgs--Doublet--Model (2HDM),
and how the presence of these three--body decays may affect the current
search techniques.
Some attention will be also given to a general model with $N\ge 3$ doublets, 
which we shall call the Multi--Higgs--Doublet--Model (MHDM).
One particular form of the 2HDM has received
substantial attention in the literature, mainly due to the fact that it is 
the structure of the minimal supersymmetric extension (MSSM) of the SM 
\cite{Gun}. However, there are four variants of
the 2HDM which differ in how the doublets are coupled to the 
fermions \cite{Bar} (we are concerned with natural flavour conservation).
In Ref.~\cite{Gross} these are referred to as Models~I, I$'$, II and II$'$,
with Model~II appropriate for the MSSM. The phenomenology of the
four models can be quite different, both in the charged and neutral
sector $\cite{Akestir} \to \cite{Akerferm}$.
 
Model~I \cite{Hab} has received relatively little attention in the 
literature,
although among other features allows the possibility of
a $H^{\pm}$ in the range of LEP2 \cite{Hew} and the phenomena known as 
"fermiophobia" $\cite{Diaz}\to \cite{Tetferm}$.
Fermiophobic Higgs bosons ($H_F$) are searched for actively at the Tevatron
\cite{Fermtev} and LEP \cite{FermLep},
using direct production methods that make use of the $ZZH_F$ coupling.
Existing limits ($M_{F}\ge 90$ GeV, $95\%$ {\sl c.l}) 
 only apply for a $H_F$ with SM strength coupling, although 
in general this coupling will be suppressed, thus allowing a 
lighter $H_F$ to be hidden. This suppression is always possible in the
general 2HDM for the lighter CP--even eigenstate $h$, and allows the
possibility of an undetected Higgs boson with $M_h\le 40$ GeV.
Such a hidden Higgs boson,
whether fermiophobic or not, could be produced by the above mentioned 
3--body decay of a heavier Higgs boson if the branching ratio (BR)
were sufficiently large. If the BR were dominant, 
present Higgs search techniques in these models would need to be changed. 
The aforementioned three--body decays have so far only appeared in the
context of the MSSM \cite{3body} and have limited importance, although we 
shall see that their strength can be considerably greater in the general 2HDM.

The paper is organized as follows. In Section 2 we introduce the
models in question and display the couplings of the Higgs bosons
to the fermions. Section 3 considers constraints on the masses
and couplings of the scalars from precision measurements. Section 4
 briefly reviews the current literature on light, hidden neutral 
Higgs bosons, while Section 5 considers the impact of the three--body
decay channels on the BRs. In Section 6 we build on the work of 
Ref.~\cite{Akestir} and consider the possibility of a 
large BR~$(H^{\pm}\to cb)$ which is only allowed in the MHDM.
Finally, Section 7 contains our conclusions.

\section{The Models}

The theoretical structure of the 2HDM is well known \cite{Gun},
while the charged Higgs sector and neutral
Higgs sectors of the MHDM have been studied in Ref.~\cite{Gross} 
and Ref.~\cite{Akeneu} respectively.
The CP conserving 2HDM which is usually considered in the literature
contains an important parameter
\begin{equation}
\tan\beta=v_2/v_1
\end{equation}
 with $v_1$ and $v_2$ being real
 vacuum expectation values (VEVs) of the two Higgs doublets, and
 $v^2=\sum_{i=1}^N v_i^2=246^2$ GeV${^2}$ for $N$ doublets.
In a MHDM it is usually assumed that one of the charged
 scalars is much lighter than the others
and thus dominates the low--energy phenomenology. 
For the charged Higgs interactions with the fermions 
 the relevant part of the Lagrangian is \cite{Gross}
\begin{equation}
{\cal L} =(2\sqrt2G_F)(X\overline{U}_LVM_DD_R+Y\overline{U}_RVM_UD_L
+Z\overline{N}_LM_EE_R)H^++h.c.
\end{equation}
Here $U_L$, $U_R$ ($D_L$, $D_R$) denote left-- and right--handed up (down)
type quark fields, $N_L$ is the left--handed neutrino field,
and $E_R$ the right--handed charged lepton field. $M_D$, $M_U$, $M_E$ are the
diagonal mass matrices of the down type quarks, up type quarks and
charged leptons respectively. $V$ is the CKM matrix.
For the four distinct versions of the 2HDM the couplings $X$, $Y$ and $Z$ are
given by the entries in Table~1 \cite{Bar}.
\begin{table}[htb]
\centering
\begin{tabular} {|c|c|c|c|c|} \hline
 & Model I & Model I$'$ & Model II & Model II$'$  \\ \hline
$X$ & $-\cot\beta$ & $-\cot\beta$ & $\tan\beta$ & $\tan\beta$ \\ \hline
$Y$ & $\cot\beta$  & $\cot\beta$ & $\cot\beta$ & $\cot\beta$ \\ \hline
$Z$ &  $-\cot\beta$  & $\tan\beta$  & $\tan\beta$  & $-\cot\beta$ \\ \hline
\end{tabular}
\caption{The values of $X$, $Y$ and $Z$ in the 2HDM}
\end{table}
In the MHDM $X$, $Y$ and $Z$ are {\it
arbitrary} complex numbers which originate from a $N\times N$ charged
scalar mass matrix. It is apparent that the models may differ
 significantly in their phenomenology.
For the couplings of the CP--odd pseudoscalar ($A$), one may use Table 1 with 
 $X$ interpreted as the coupling to $d\overline d$,
$Y$ the coupling to $u\overline u$ and $Z$ the coupling to $l\overline l$.
For the lighter CP--even eigenstate $h$ one finds the values given
in Table 2, with $\alpha$ a mixing angle in the CP--even sector.
\begin{table}[htb]
\centering
\begin{tabular} {|c|c|c|c|c|} \hline
& Model~I & Model~I$'$ & Model~II & Model~II$'$  \\ \hline
$hu\overline u$ & $\cos\alpha/\sin\beta$ & $\cos\alpha/\sin\beta$
& $\cos\alpha/\sin\beta$  & $\cos\alpha/\sin\beta$      \\ \hline
$hd\overline d$ & $\cos\alpha/\sin\beta$& $\cos\alpha/\sin\beta$ 
&$-\sin\alpha/\cos\beta$ &$-\sin\alpha/\cos\beta$      \\ \hline
$hl\overline l$ &  $\cos\alpha/\sin\beta$  & $-\sin\alpha/\cos\beta$ 
  &$-\sin\alpha/\cos\beta$ 
&  $\cos\alpha/\sin\beta$\\ \hline
\end{tabular}
\caption{The fermion couplings of $h$ relative to those for the
 minimal SM Higgs boson ($\phi^0$).}
\end{table}
For the heavier CP--even $H$ one must make the replacements 
$\cos\alpha\to \sin\alpha$ and
$-\sin\alpha\to \cos\alpha$ in Table 2.

\section{Constraints and Branching Ratios}

Precision measurements of the process $b\to s\gamma$ impose the
severest constraints on the mass of the
charged scalar of the 2HDM (Models~II and II$'$). For a general review of how 
new physics affects this decay see Ref.~\cite{Hew}.
The diagrams which contribute to this process are essentially
the same as those for the
SM with the $W^{\pm}$ replaced by $H^{\pm}$.

The CLEO collaboration obtained the value \cite{CLEO}
\begin{equation}
{\rm BR} (b\to s\gamma)= (2.32\pm 0.51\pm 0.29 \pm 0.32)\times 10^{-4},
\end{equation}
and more recently ALEPH \cite{Aleph} have found 
\begin{equation}  
{\rm BR} (b\to s\gamma)= (3.38 \pm 0.74 \pm 0.85)\times 10^{-4}.
\end{equation} 
It is known that for $H^{\pm}$ of Model~I, I$'$ and MHDM  
one cannot obtain a mass
bound independent of $X$ and $Y$ (e.g. see Refs.~\cite{Gross},
\cite{Akestir}).  For $H^{\pm}$ of 
Model~II and II$'$ one finds the constraint $M_{H^\pm}\ge 330$ GeV 
for any value of $\tan\beta$ \cite{Giudice}.
Measurements of $b\to s\gamma$ do constrain $\tan\beta$ in Model~I and I$'$ 
for a given Higgs mass
although  the decay $Z\to b\overline b$ imposes stronger
constraints. Ref.~\cite{Giudice}
shows that from the latest $R_b$ measurements
one can obtain the 
 respective bounds of $\tan\beta\ge 1.8, 1.4, 1.0$ ($95\%$ c.l) for
$M_{H^\pm}=85, 200, 425$ GeV. 
 The BRs for $H^{\pm}$ of mass 80 GeV are given in Table 3, excluding the
possibility of three--body decays. The Higgs mass determines the energy 
scale of the decay and so one must evaluate the quark masses at the 
scale $Q=M_{H^\pm}$, and 
these BRs improve those that we gave in Ref.~\cite{Akestir} (which were 
purely tree--level). Note that in Model~I the BRs are independent 
of $\tan\beta$, while in Model~I$'$ there is a $\tan\beta$ dependence
which causes the inequalities in Table 3 (we take $\tan\beta\ge 1.8$).
For the MHDM it is not possible to predict the BRs since the parameters
$X$, $Y$, and $Z$ may be varied independently of each other.
\begin{table}[htb]
\centering
\begin{tabular} {|l|c|c|c|} \hline
 & cs & cb & $\tau\nu_{\tau}$  \\ \hline
 $H^{\pm}$ (Model I) & $34.03\%$  & $1.22\%$ & $64.75\%$  \\ \hline
 $H^{\pm}$ (Model I$'$) & $\le 4.76\%$ & $\le 0.17\%$ & $\ge 95.07\%$ \\ \hline
 $H^{\pm}$ (MHDM)& $0\%\to 100\%$ & $0\%\to 100\%$ &  $0\%\to 100\%$ \\ \hline 
\end{tabular}
\caption{The branching ratios of $H^{\pm}$ excluding 3--body decays.}
\end{table}
 The $cb$ channel is of order one percent in Models~I and I$'$ 
 due to heavy CKM 
matrix suppression, although in a MHDM it is
possible to enhance this channel (\cite{Akestir} and Section 7).  

For all the charged scalars that we consider there exists an 
experimental lower bound from LEP of $54.5$~GeV \cite{Cha} which assumes
$cs$ and $\tau\nu_{\tau}$ decays of $H^\pm$. This limit may not be valid 
if three--body decays dominate, although there exists a decay mode 
independent bound $M_{H^\pm}\ge 40$ GeV
from considering visible decays of new particles contributing to the
$Z$ width \cite{monig}.
For the neutral sector there is no limit on $M_A$ since its standard
production mechanism is in association with $h$, and so if $M_h$ is 
sufficiently large this channel would not be
open. The literature mentions other ways of producing a light $A$ which
we shall briefly review in Section 4.
For $h$ the process $e^+e^-\to Z^{(*)}\to hf\overline f$ is 
available although this production method  
is proportional to $\sin^2(\beta-\alpha)$ and so may be suppressed.
 Hence for small values of 
$\sin^2(\beta-\alpha)$ a light $h$ with mass significantly lower than
that of the current SM bound ($M_{\phi^0}\ge 87.6$ GeV)
\cite{moriond} is not ruled out.
We note that in the MSSM it is possible to put actual limits of 
$M_h\ge 70.7$ GeV (all $\tan\beta$) and
$M_A\ge 71.0$ GeV ($\tan\beta\ge 1$) \cite{HAlim}.
For the heavier CP--even scalar ($H$) the presence of a small 
$\sin^2(\beta-\alpha)$ would automatically force $\cos^2(\beta-\alpha)
\to 1$, thus enabling it to be produced with almost $\phi^0$ strength
rate. However, in this scenario 
the decay $H\to hh$ may be open in all models and could dominate the
standard $f\overline f$ decays.
Assuming production via $e^+e^-\to Z^*\to HZ$, the signature
would be a final state of 6 fermions, with four of them likely
to be $b$ quarks coming from $h\to b\overline b$ decay. 
Such an event signature would be very similar to that coming from the
 process $e^+e^-\to Zh$ with subsequent decay $h\to AA$, 
 which is considered in the searches, and so the current analysis
 should still be applicable to $H$.

We do not believe that $\rho$ parameter constraints in these scenarios
of light Higgs bosons have been considered, especially bearing in 
mind the current bound on $M_{H^\pm}\ge 330$ GeV for the
charged Higgs of Model~II.  One defines $\rho^0$ as: 
\begin{equation}
\rho^0={M^2_W\over \rho M_Z^2\cos^2\theta_W}
\end{equation}
Here $\rho$ in the denominator contains all purely SM radiative
corrections, while $\rho^0 \equiv 1$ in the absence of new physics.
 In the 2HDM there are extra contributions to $\rho^0$ \cite{Hollik}
given by:
\begin{equation}
\Delta\rho^0={G_F\over 8\pi^2\sqrt 2}\left[\sin^2(\alpha-\beta)F
(M^2_{H^{\pm}},\,M^2_{A},\,M^2_{H})+\cos^2(\alpha-\beta)
F(M^2_{H^{\pm}},\,M^2_{A},\,M^2_{h})\right]\,,
\end{equation}
with 
\begin{equation}
F(a,\,b,\,c)=a-{bc\over b-c}{\rm ln}{b\over c}-{ab\over a-b}{\rm ln}
{a\over b}-{ac\over a-c}{\rm ln}{a\over c }\,. 
\end{equation}
Ref. \cite{Langacker} shows that $-0.0017\le \Delta\rho^0\le
0.0027$ at the $2\sigma$ level.
For the case of a light $h$ one requires 
$\sin^2(\beta-\alpha)\to 0$, and so the dependence on the heavier
neutral CP--even scalar $(H)$ drops out. If we demand a light $A$  no
condition on $\sin^2(\beta-\alpha)$ is required, and so
 $M_H$ cannot be neglected from Eq. (5).
Therefore the case of a light $h$ allows the above formula to simplify
 considerably and in Fig. 1 we plot the contribution $\Delta\rho^0$ against 
 $M_A$ for three values of $M_{H^\pm}$, fixing $M_h=20$ GeV. 
  We see from Fig. 1 that to maintain $\Delta\rho^0$ within the $2\sigma$
limits (denoted by the horizontal lines) one requires $M_A\ge 250$
 GeV in Model~II, assuming $M_{H^\pm}\ge 330$ GeV.\footnote{On completion of
 this work we became aware of the new bound $M_{H^\pm}\ge 165$ GeV \cite{bsy},
 \cite{sola}.
 This would eliminate the requirement $M_A\ge 250$ GeV.}
In Models~I and I$'$ since ${H^\pm}$ may be light 
 one may simultaneously allow a light $A$ and $h$ without violating the
$2\sigma$ limits.

\begin{figure}[htb]
\centerline{\protect\hbox{\psfig{file=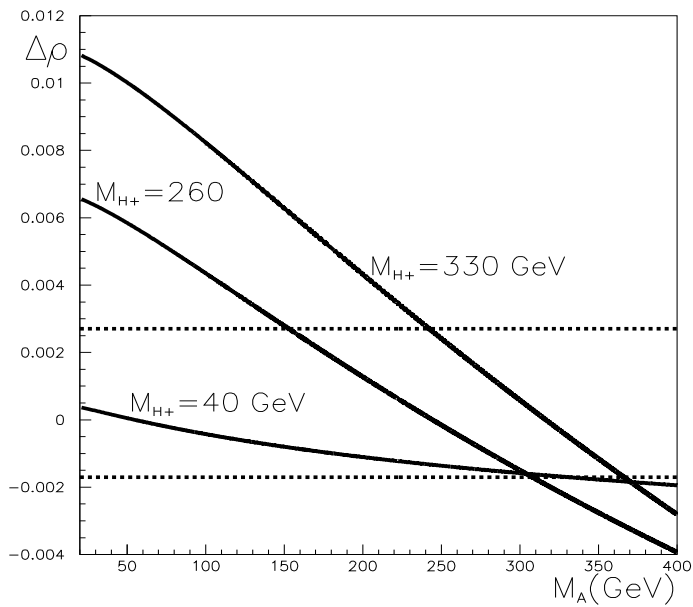,height=10cm,width=10cm}}}
\caption{$\Delta\rho^0$ as a function of $M_A$ with $M_h=20$ GeV.}
\label{cha1}
\end{figure}

\section{A light $h$ or $A$ at LEP2}
In this section we briefly review the current status of the
literature on a light $A$ or $h$ i.e. the case of $M_h$ or $M_A\le 40$ GeV.
It is usually
assumed that the sum of $M_A$ and $M_h$ is greater than $M_Z$, since the
excluded region in the $M_A,M_h$ plane has the form $M_A+M_h\ge 90\to 110$
GeV \cite{HAlim}. We note that the searches  
do not consider the possibility of three--body decays to a lighter Higgs,
 although if such decays were present with a large BR they would
 have given a similar 6 fermion signature to those
 already searched for at LEP (see Section 5.3). 
 Therefore we shall assume that $M_A+M_h\ge 90\to 110$ GeV.
 
 A light $h$ or $A$ has really only been considered in the context of
 the 2HDM (Model~II) and has received limited attention 
 $\cite{Mar}\to \cite{Keith}$. These papers do not consider the 
 three--body decay 
  of a heavier Higgs boson to a light $h$ or $A$ plus a virtual vector boson,  
  and this would be
 an alternative way of producing a $h$ or $A$ which has thus far 
 escaped  detection.  We shall give particular attention to the 2HDM 
 Model~I and I$'$, whose Higgs bosons may possess a significantly
 different phenomenology to those of Model~II.  
 In the following paragraphs we briefly summarize the existing 
 methods of producing a light
 $h$ and $A$, and check to see if they are relevant for Models~I and I$'$.
 
 Ref.~\cite{Kraw} studied the Yukawa production method, $e^+e^-\to
 b\overline bh(A)$.  This process may be important in Model~II 
 in the case of large $\tan\beta$, since the $h(A)b\overline b$
 coupling is proportional to $\tan\beta$. In Model~I$'$ the 
 $h(A)b\overline b$ coupling is proportional to $\cot\beta$ and so one
 would have to rely on $h(A)$ emission from $\tau\tau$ pair production.
 This would give a much lower rate since the coupling $h(A)\tau\tau$ in
 Model~I$'$ is smaller than $h(A)b\overline b$ in Model~II by a factor 
 $3m^2_b/m^2_{\tau}$.
 In Model~I all the fermion couplings scale as $\cot\beta$ and the Yukawa 
 method would not be effective. 

Ref.~\cite{Mar}  considers the process $e^+e^-\to Z\to hhf\overline f$, whose
 dominant contribution comes from production of either $H$ or $H^*$ with
 subsequent decay to $hh$. If H is on--shell then the rate can be large and
 is therefore a process in the same spirit to the methods we
  consider here -- that
 is, producing $h(A)$ by the decay of an on--shell Higgs particle. This method
 was only considered for Model~II, although it may also be used for
  Model~I and I$'$. The $Hhh$ coupling is model 
  independent, since this coupling originates from the Higgs potential.
   In the case of $H$ being on--shell,
 the BR$(H\to hh)$ will be different depending on the model in question.
  Applying this method to the case of a light fermiophobic
 Higgs (i.e. Model~I with $\cos\alpha\to \pi/2$ and $\sin^2(\beta-\alpha)$ 
 small) would cause the fermion couplings
 of $H$ to be scaled by a factor $1/\sin\beta$ relative to the SM Higgs.
 The production channel $e^+e^-\to Z^{(*)}\to 
 HZ$ would then proceed with a rate close to that of the SM Higgs, 
 and the subsequent decay $H\to H_FH_F$ would  
 give the signature $\gamma\gamma\gamma\gamma$ and $f\overline f$ in opposite
 hemispheres. This topology
 would pass the current selection criteria for a $H_F$ which
 demands 2 isolated $\gamma$ recoiling against a fermion pair \cite{FermLep}. 
 In the case of $H$ being off--shell we still have a model dependent rate
  since the width of $H$ will appear explicitly in the propagator.  
 
 Refs.~\cite{Chang} and \cite{Keith} consider the production of $AAA$ 
via $e^+e^-\to Ah^*\to AAA$ in the case when an on--shell $h$ cannot be
 produced. This process will again be model dependent for the reasons
 cited above. The case when the $h$ is on--shell is considered in the 
 searches in Ref.~\cite{HAlim}, and we shall be covering this decay
 in more detail in Section 5.3. 
 
\section{Three--Body decays of Higgs Bosons}
In the following sections $5.1\to 5.5$ we study the impact 
of the three--body decays
on the BRs and searches for $H^{\pm}$, $h$, and $A$ in the context of the 
2HDM. We shall see that their importance  varies from model to model 
and can be especially significant in Model~I since $H^{\pm}$ and $A$ 
 can decouple from the fermions at large $\tan\beta$
 (see Table 1). 
We stress that an important use of these 3--body decays is that they
may enable the detection of a light Higgs particle which has eluded 
current searches at LEP1 due to suppressed couplings to the $Z$. 

\subsection{The decays $H^{\pm}\to W^{*}h$ and $H^{\pm}\to W^{*}A$}
In this subsection we consider the three--body decay 
$H^{\pm}\to h(A)f'\overline f$ mediated by a virtual $W^{\pm}$. 
Such a decay is not possible in the 2HDM (Model~II) at LEP2 due to the bound 
$M_{H^\pm}\ge 330$ GeV. In Model~I and I$'$ one may avoid the
mass constraints from $b\to s\gamma$, as explained in Section 3, and so these
models may possess a charged Higgs in the discovery range of LEP2.  
Studying the BR of this channel is important for two reasons:\\
\begin{itemize} 
\item[{(i)}] It may vastly alter current charged Higgs searches at LEP2, 
which always assume decays to $\tau\nu_{\tau}$ and $cs$.
\item[{(ii)}] It would be an alternative way of discovering
a light $h$ or $A$ which is escaping current searches due to weak
couplings to $Z$. 
\end{itemize}
If the three--body decay channel for $H^{\pm}$
were dominant it would invalidate the current limit $M_{H^\pm}\ge 54.5$ GeV
\cite{Cha},
and justify the use of the weaker limit $M_{H^\pm}\ge 40$ GeV \cite{monig}.  
Point (ii) is of particular interest for a fermiophobic Higgs which is searched
for actively at the Tevatron and LEP, and may be hidden due to suppressed
couplings to vector bosons. LEP \cite{FermLep}
 uses the standard Bjorken process, while
the Tevatron \cite{Fermtev} uses $q'\overline q\to H_FW$, which also depends 
on the vector boson
coupling. In contrast, the cross--section $e^+e^-\to H^+H^-$ does not suffer
mixing angle suppressions \cite{Kom}, instead being
dictated by $M_{H^\pm}$. If BR$(H^{\pm}\to W^*h, (A)$)
 were large then it would be a copious source of a light $H_F$.
We stress that this detection channel channel is not possible at LEP2 
for a light $h$ and $A$ of the 2HDM (Model~II) considered
by Refs.~$\cite{Mar}\to \cite{Keith}$ since $M_{H^\pm}\ge 330$ GeV. 
\begin{figure}[htb]
 \centerline{\protect\hbox{\psfig{file=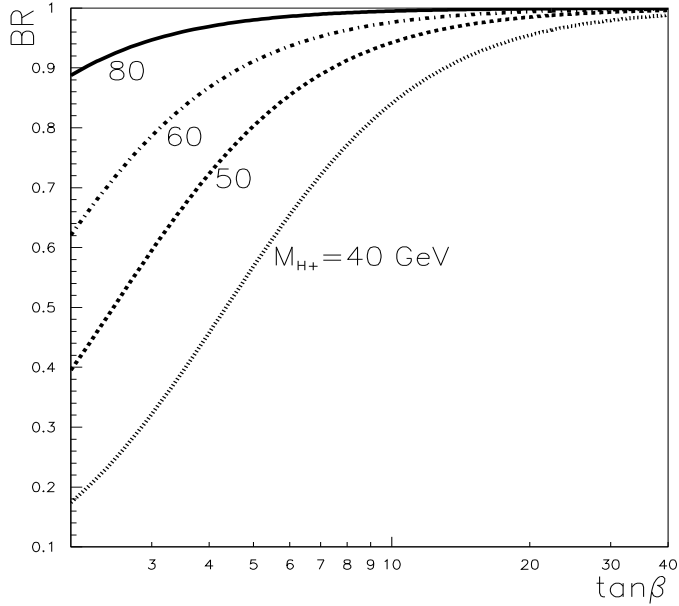,height=10cm,width=10cm}}}
 \caption{BR$(H^\pm\to W^*h(A))$ in Model~I 
 against $\tan\beta$ for $M_{h(A)}=10$ GeV.}
 \label{cha2}
 \end{figure}
The partial widths for this channel are as follows \cite{3body}, \cite{Arib}:
\begin{equation}
\Gamma(H^{\pm}\to hW^{\pm*}\to hf\overline f')={9G^2_FM^4_W\over 
16\pi^3}\cos^2(\beta-\alpha)M_{H^\pm}G_{hW}\,.
\end{equation}
\begin{equation}
\Gamma(H^{\pm}\to AW^{\pm*}\to Af\overline f')={9G^2_FM^4_W\over 
16\pi^3}M_{H^\pm}G_{AW}\,.
\end{equation}
The functions $G_{AW}$ and $G_{hW}$ depend on the masses
of the particles in the decay and are given in Refs.~\cite{3body}
and \cite{LEP}.
 We note that the function $G_{ij}$ displayed in Refs.~\cite{3body}
 and \cite{LEP} (where $i$ and $j$ refer to a Higgs and a vector boson)  
 contains a typing error, which is corrected by changing 
 the final term $-2\lambda_{ij}/\kappa_j$
 to $+2\lambda_{ij}/\kappa_j$. The function $G_{ij}$ is an approximation
 of the numerical integration over the Dalitz plot,
 but breaks down in the parameter space of interest to us 
 ($M_{H^\pm}\le M_W$). 
Therefore in the analysis that follows we shall evaluate $G_{ij}$
numerically. In the case of Eq. (8) the condition for a light
$h$ causes 
$\cos^2(\beta-\alpha)\to 1$ and so enhances this width. 
  Since we are interested in the case of a light $h$ 
 we shall take $\cos^2(\beta-\alpha)\approx 1$, and
 so the results for the decay to $h$ and $A$ are more or less identical.
 We now consider in turn Models~I and I$'$.
 \begin{figure}[bt]
 \centerline{\protect\hbox{\psfig{file=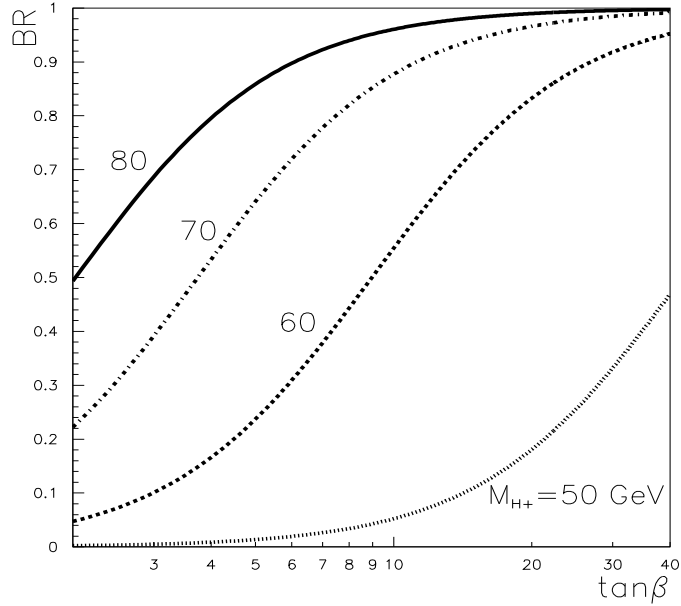,height=10cm,width=10cm}}}
 \caption{Same as Fig. 2 but for $M_{h(A)}=40$ GeV.}
 \label{cha3}
 \end{figure} 
  
 \subsubsection{Model I}
 Fig. 2 displays BR($H^{\pm}\to W^{*}h(A))$ in Model~I as a function
 of $\tan\beta$ for 4 different values of $M_{H^\pm}$
 ($40\to 80$ GeV), with $M_h(M_A)=10$ GeV. As one can see, the BR of the
 three--body decay is close to $100\%$ over the majority of the
 $\tan\beta$ parameter space. The difference in mass
 between $M_{H^\pm}$ and $M_h$($M_A$) is important since it determines how
 off--shell the vector boson is, and so the curves for lower $M_{H^\pm}$ take
 longer to reach $\approx 100\%$. In Fig. 3 we take $M_h(M_A)=40$ GeV.  
 We conclude that the three--body decays, if open, can be of great
 importance in Model~I since the standard decays to 2 fermions are
 proportional to $\cot^2\beta$. In addition, the possible large BR of this
 channel may have allowed $H^{\pm}$ to have avoided previous
 searches, and thus the limit $M_{H^\pm}\ge 40$ GeV should only be applied.
 The principal decays of $h$ and $A$ 
  would be to $b\overline b$, unless $h$ is fermiophobic,
  in which case $H_F\to \gamma\gamma$ would dominate.
 In the case of fermiophobia  one could find  
 $\gamma\gamma f\overline f$ events in each hemisphere. LEP
  currently searches
  for $\gamma\gamma$ recoiling against 
  $f\overline f$ \cite{FermLep}, demanding isolated photons. Although in our
  new signature each photon pair would be accompanied by a pair of quarks,
  they would still pass the current selection criteria -- in fact the
  efficiency of detecting any two of the four photons would increase
  by a factor $\approx 1.2$ \cite{turcot} relative to the efficiency for the
  $e^+e^-\to H_FZ\to \gamma\gamma f\overline f$ channel.
  We conclude that a very light $H_F$ could
 be copiously produced in $H^\pm$ decays and it should be possible
  from the topology
 to see whether one has registered a signal in the 
 $e^+e^-\to H_FZ$ channel or $H_FH_FW^*W^*$ channel. The best discriminator
 would be the detection of 3 or more of the photons, which has a very
 small SM background and would have an efficiency $\approx 1.05$ times
  that for the $e^+e^-\to H_FZ$ channel.  
 We stress that a light $H_F$ may continue avoiding searches in the
 $e^+e^-\to H_FZ$ channel due to weak coupling to the $Z$,
  although could be detected via charged Higgs
 decays over a wide range of $\tan\beta$ values, provided there are
 enough pair produced charged Higgs bosons. Lack of signal 
 could be used to rule out regions of parameter space for
 $M_F$/$M_{H^{\pm}}$/$\tan\beta$.
 
 As we pointed out in
 Ref. \cite{Aketrip} an additional
 condition on the existence of a light $H_F$ is   
 $M_{F}+M_A\ge 160$ GeV. This is due to the fact that the channel
 $e^+e^-\to AH_F\to \gamma\gamma f\overline f$ is also searched for in Ref.
 \cite{FermLep}
 and is complementary to $e^+e^-\to H_FZ\to \gamma\gamma f\overline f$
 Therefore the former must be closed kinematically 
 if one wishes to consider suppressed
 $e^+e^-\to H_FZ$ production and the possibility of a light $H_F$.
 
 If $h$ is not fermiophobic then one could have final states of 8 fermions,
 with four of them likely to be $b$ quarks. Such a signature might
 allow detection of $H^{\pm}$ in the difficult $M_{H^\pm}\approx M_W$ 
 region, although a full analysis is beyond the scope of this paper.
 
 \subsubsection{Model I$'$}
 
 In Model~I$'$ the three--body decay can only be
 strong only at small $\tan\beta$, 
  which we can see from Fig. 4.
 This is because the $H^{\pm}\to \tau\nu_{\tau}$ decay width is proportional
 to $\tan^2\beta$ and so $H^{\pm}$ does not decouple from the fermions
 as $\tan\beta$ increases. Regarding the possible signatures,
  $h(A)\to \tau\tau$ and $h(A)\to b\overline b$ would have
  equal rates for $\tan\beta\approx 2$,
  with the former quickly approaching $100\%$ as $\tan\beta$ increases.
  Hence one would expect a final state of 8 fermions, with 4 of them likely
  to be $\tau$ leptons. Note that
  we are interested in the region $\alpha\approx \beta$ in order 
  to ensure a light $h$, and so the $h\to \tau\tau,b\overline b$ couplings
  lose their
 $\alpha$ dependence and depend purely on $\tan\beta$. 
  \begin{figure}[htb]
 \centerline{\protect\hbox{\psfig{file=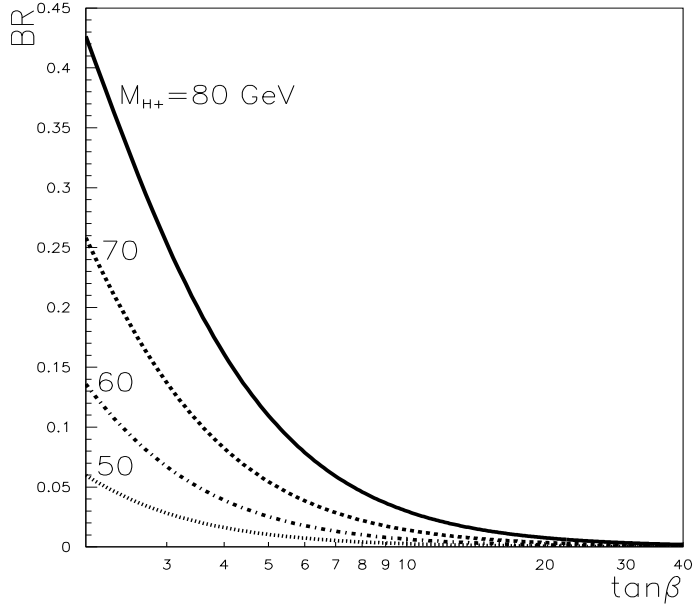,height=10cm,width=10cm}}}
 \caption{Same as Fig. 2 but for Model I$'$.}
 \label{cha1}
 \end{figure}
  
 \subsection{The decay $h\to H^{\pm}W^*$}  
  For the decay $h\to H^{\pm}W^*$ one can only consider Models~I 
  and I$'$, since in Models~II and II$'$ the bound $M_{H^\pm}\ge 330$ GeV
  disallows this process at LEP2.   
  The production mechanisms for $h$ are
  $e^+e^-\to Z^*\to hZ$ and (if open) $e^+e^-\to Z^*\to Ah$. These
  two channels are complementary, the former being proportional
  to $\sin^2(\beta-\alpha)$ and the latter to $\cos^2(\beta-\alpha)$.
  
 \begin{figure}[htb]
 \centerline{\protect\hbox{\psfig{file=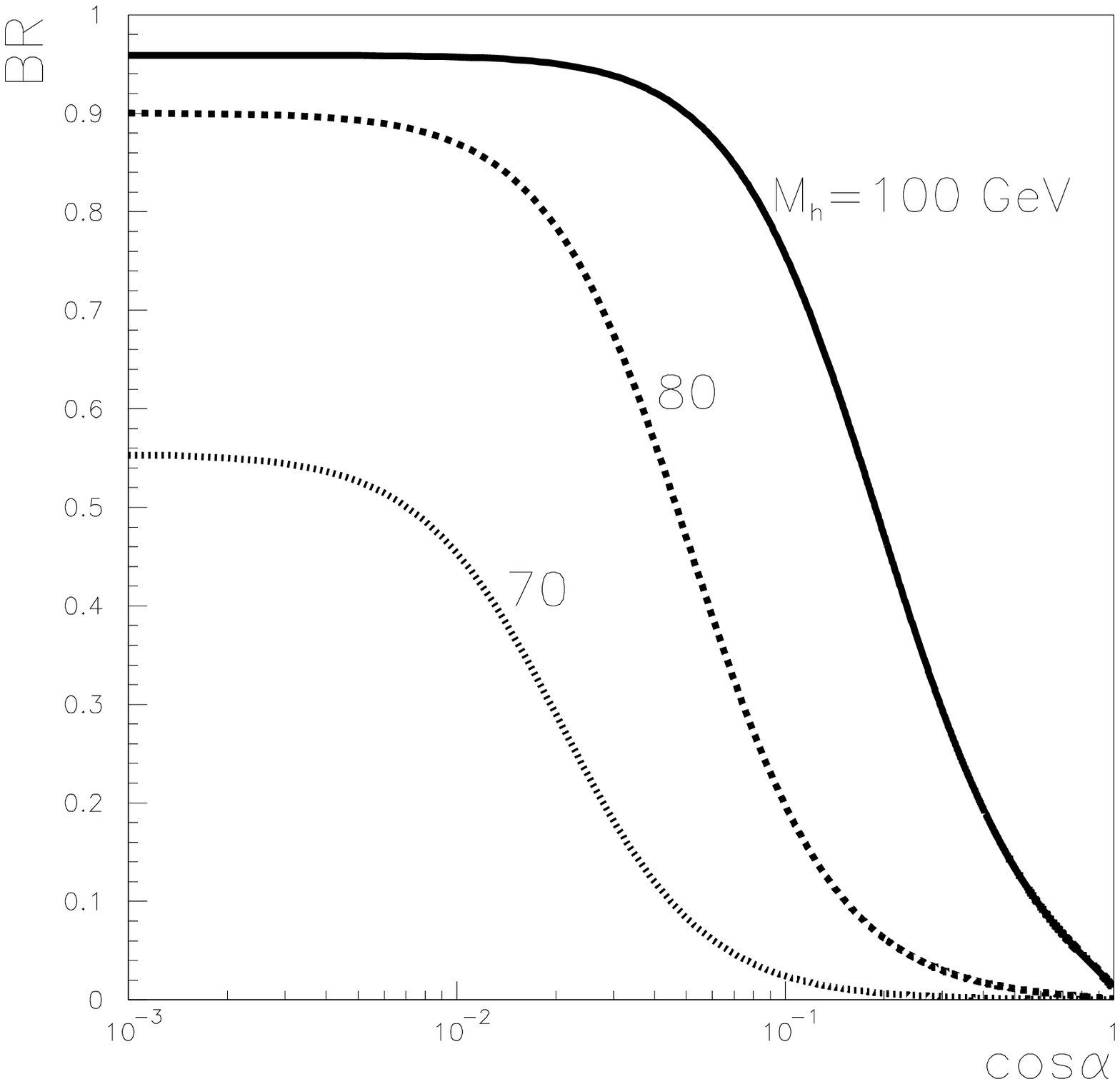,height=10cm,width=10cm}}}
 \caption{BR$(h\to H^{\pm}W^*$) against $\cos\alpha$ for Model I with
 $M_{H^{\pm}}=55$ GeV.}
 \label{cha8}
 \end{figure} 
  \begin{figure}[htb]
 \centerline{\protect\hbox{\psfig{file=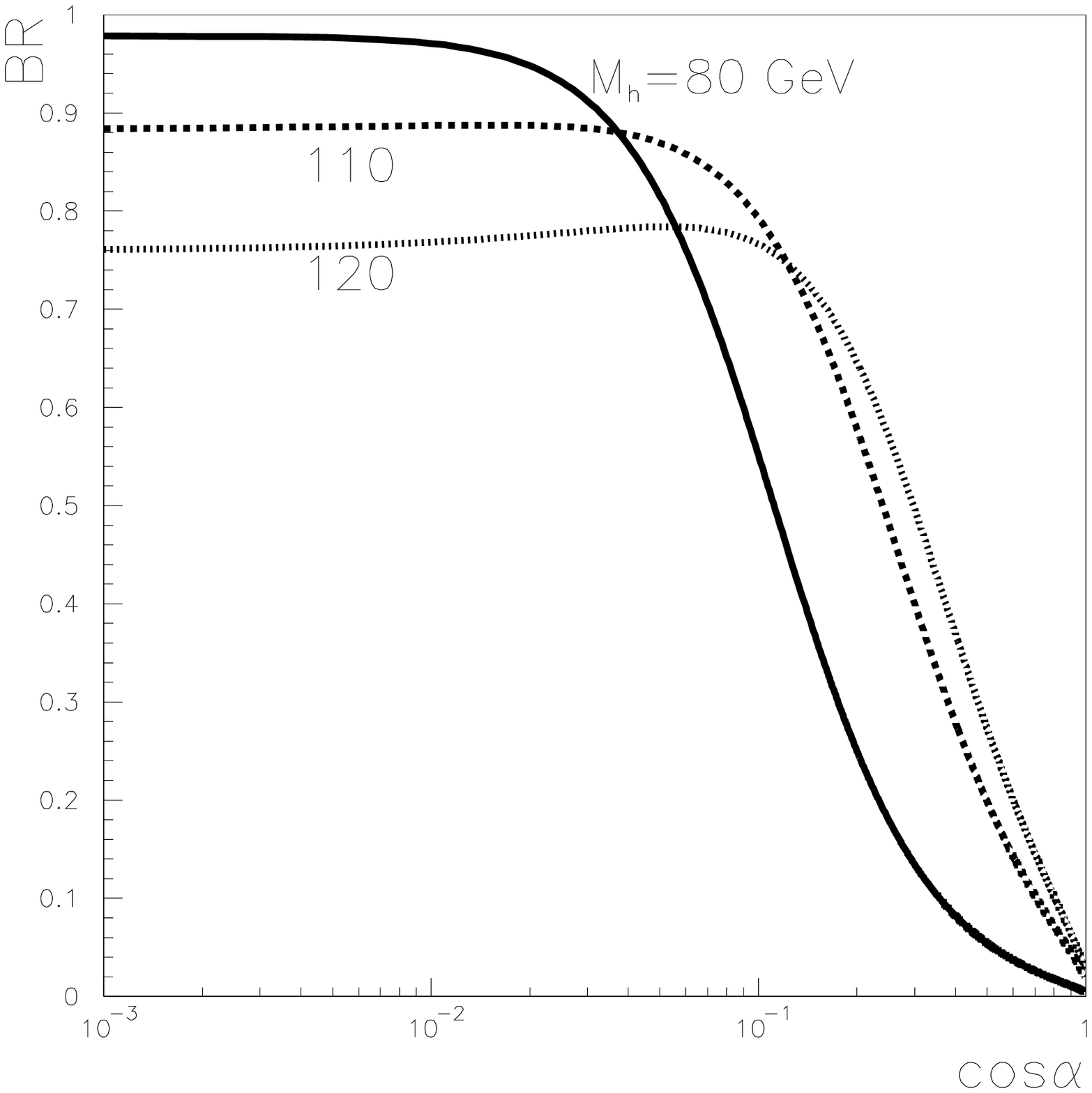,height=10cm,width=10cm}}}
 \caption{BR$(h\to AZ^*)$ against $\cos\alpha$ for Model I
  when $h\to AA$ is not allowed.}
 \label{cha8}
 \end{figure} 
  The couplings of $h$ to $f\overline f$ involve the
  mixing angle $\alpha$ (see Table 1), and in Model~I have a minor 
  dependence on $\tan\beta$; this is because the factor 
  ($\sin\beta)^{-1}$
  appears explicitly and for $\tan\beta\ge 2$ takes values between
  1 and 1.12. In Fig. 5 (for Model~I) we plot BR$(h\to H^{\pm}W^*)$ as
   a function of 
  $\cos\alpha$, fixing $\tan\beta=2$, for values of $M_h$ up to 100 GeV
  (we take $M_{H^{\pm}}=55$ GeV). From the
  figure we see that the BR for the three--body decay is 
  at a maximum at $\alpha=\pi/2$,
  (the condition for fermiophobia), and for low values of
  $\cos\alpha$ the competing decays are $h\to \gamma\gamma$ and $h\to WW^*$.
  As $\cos\alpha$ increases the standard decays $h\to f\overline f$ start to
  gain in strength.
  
  In Model~I$'$ there is a strong $\tan\beta$ dependence, since the 
  coupling $h\to \tau\tau$ grows as $\tan\beta$. We find that 
  BR$(h\to H^{\pm}W^*)$ peaks at $\approx 5\%$($\approx 0.5\%$) for 
  $M_h=100$ GeV with $\tan\beta=2$(10). Regarding the possible signatures
  in both models, large multiplicity fermion events would be possible.
  
  \subsection{The decay $h\to Z^*A$}
  We now consider the case of $M_A\le M_h$ which
  allows the decay $h\to AZ^*$, and we shall assume that
  decays to $H^{\pm}W^*$ are not open. If $M_h\ge 2M_A$ the channel
  $h\to AA$ would be open as well. The experimental
  signature of these two decays would be similar, both giving  6 fermion 
  final states coming from either production mechanism.
  There are four different topologies here which we list in Table 4,
  and the colon separates the particles into hemispheres.
\begin{table}[htb]
\centering
\begin{tabular} {|l|c|c|} \hline
& $h\to AZ^*$   & $h\to AA$  \\ \hline
$e^+e^- \to hZ$ & $Z:Af\overline f$  & $Z:AA$ \\ \hline
$e^+e^- \to hA$  & $Af\overline f:A$   & $A:AA$ \\ \hline
\end{tabular}
\caption{The 6 fermion topologies originating from $h$ and/or $A$
production.}
\end{table}
   We note that in principle these 4 distinct topologies could be 
  distinguished by measuring the invariant masses of the jets and/or using
  $b$--tagging techniques. 
  The current searches do consider
  6 fermion final states, but from the lack of signal rule out regions
  in the $M_h-M_A$ plane only for $M_h\ge 2M_A$. For the region $M_h\le 2M_A$
  only the decays $h\to f\overline f$ are considered. 
  We wish to see if the three--body decays (and thus the 6 fermion
  signature) can be important even when
  the decay $h\to AA$ is not open. In the following subsections we 
  consider the strength of the decay $h\to AZ^*$ in Models~I, I$'$ and II
  respectively. 
  \subsubsection{Model I}
  In Fig. 6 (again for $\tan\beta=2$)
   we plot BR$(h\to AZ^*)$ in Model~I for values of $M_A$ which do 
  not permit $h\to AA$ (we take $M_h=2M_A-5$ GeV). If 
   $h\to AA$ is open, it will contribute to the 6 fermion signature
   although giving a different topology as shown in Table~4.
  Note that the production process $e^+e^-\to Z^*\to hZ$ would 
   not be open for the values of $M_h\ge 100$ GeV displayed
   in Table~4. We see from
  Fig. 6 that lighter values of $M_h$ allow larger BR$(h\to AZ^*)$
    at small values
  of $\cos\alpha$ than for heavier $M_h$; as $\cos\alpha$ increases
  the curves for lighter $M_h$ fall more rapidly. 
  This can be explained as follows. Larger $M_h$ 
  increases the partial width for the three--body decay (since the
  $Z$ will be less off--shell), but also enhances
  the decay $h\to WW^*$. Hence in the small $\cos\alpha$ 
  region the curves with larger $M_h$ take lower BR values since in this 
  region the competing decay
  is $h\to WW^*$, with the fermions decoupled. 
  As $\cos\alpha$ increases the fermion decays start to
  gain in importance, and so all the curves fall to small values;
  those curves with larger $M_h$ do not fall as sharply since their 
  partial width is larger.   
  Of course, one may choose $M_h$ and $M_A$ such that
  the on--shell decay $h\to ZA$ is open. For these choices of
  Higgs masses the decay $h\to AA$ would always be open 
  (since $2M_A\le M_Z+M_A$) and so one would expect the 6 fermion signature
  to dominate the 4 fermion signature. 
  \subsubsection{Model I$'$ and Model II}
  In both these models the fermions never decouple completely from 
  $h$, and so we expect lower maximum BRs for the three--body decay. 
  In addition there will be a strong $\tan\beta$ dependence.
  In Model~I$'$ we find a peak
  BR$(h\to AZ^*)\approx 10\%$ ($1\%$) for $\tan\beta=2(10)$, with
  $M_h=120$ GeV. In Model~II these values drop to $\approx 5\%$ and 
  $\approx 0.2\%$ respectively.

\subsection{The decay $A\to Z^*h$}

These decays have not been considered so far in the searches for $A$
of the general 2HDM, and if open may
be of great importance in Model~I.  The phenomenology of $A$ in 
Model~I and Model~II has been
considered in the context of the LHC \cite{kominis},
 but this work did not consider three--body decays, and was concerned with a 
 more massive $A$ which could decay to on--shell $hZ$.
 \begin{figure}[htb]
 \centerline{\protect\hbox{\psfig{file=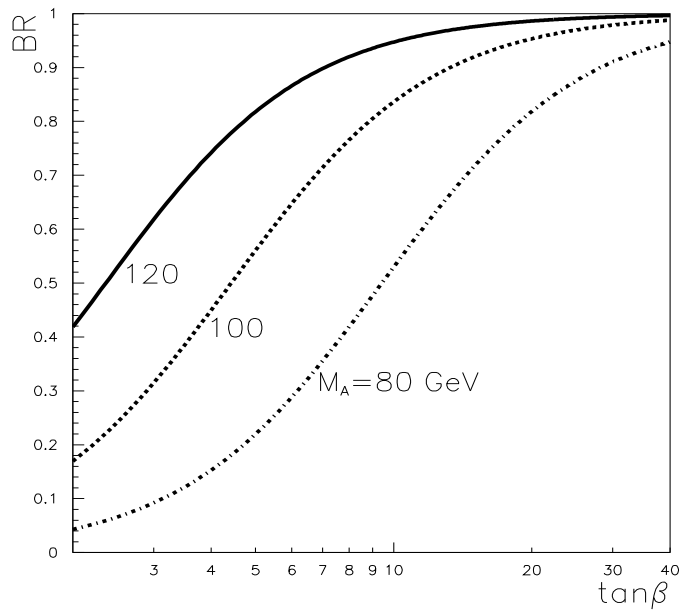,height=10cm,width=10cm}}}
 \caption{BR$(A\to Z^*h)$ in Model I against $\tan\beta$ for
 $M_h=40$ GeV.}
 \label{cha5}
 \end{figure}
The condition for a light $h$ 
causes $\cos^2(\beta-\alpha)\to 1$, and so in this case the production
cross-section $e^+e^-\to Z^*\to Ah$ would be not mixing angle suppressed.
We now consider the strength of the decay $(A\to Z^*h)$ in Models~I, I$'$ 
and II. 
\begin{figure}[htb]
\centerline{\protect\hbox{\psfig{file=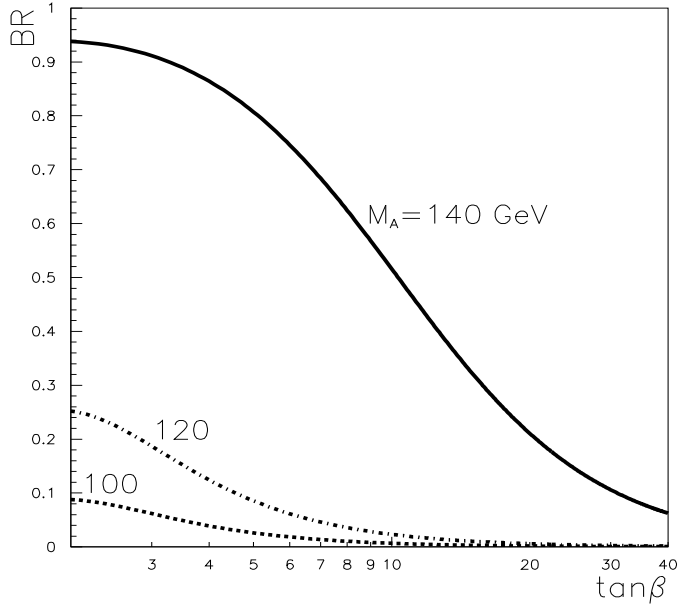,height=10cm,width=10cm}}}
 \caption{Same as Fig. 7 but for Model I$'$.}
 \label{cha6}
 \end{figure}
\subsubsection{Model I} 
In Fig.~7 we plot BR$(A\to Z^*h)$ for 3
values of $M_A$, fixing 
$M_h=40$ GeV. We require $M_A+M_h\ge 110$ GeV (from LEP1 searches), 
and deliberately consider
mass choices which make $Z$ off--shell in order to
emphasize that the off--shell decays can be prominent. 
We stress that lower values of $M_h$ would imply that the on--shell decay
$A\to hZ$ is open  and 
so the three--body BR would be $\approx 100\%$ over all the 
$\tan\beta$ range -- for this
reason we do not plot a graph. We also assume in Fig. 7 that $M_{H^{\pm}}$
is not light enough to cause a competing three--body decay. 
From Fig.~7 one can see the importance of the decay $A\to Z^*h$ when open, 
and it would be an alternative way of producing a light $H_F$ with a good
rate. The signature would be similar to that discussed in
Section 5.1.1, although in this case there would be less jets. 

Comparing Fig. 7 with Fig. 3 one sees a significant difference in the
strength of  
the respective lines for $M_{H^\pm}=80$ GeV and $M_A=80$ GeV, the former
starting at BR$\approx 50\%$ and the latter at BR$\approx 5\%$.
There
are two reasons why the three--body decay for $H^\pm$ is stronger
than that for $A$, for identical masses of the Higgs bosons in the decay,
despite the fact that the $Z$ mediated decay has a slightly stronger coupling
by a factor of $\approx 4/3$.
Firstly, the fact that $M_Z\ge M_W$ means that 
$Z$ would be more off--shell in the decay $A\to hZ^*$ than $W$
in the decay $H^\pm\to W^*h(A)$. Secondly, and more importantly, 
BR$(H^\pm\to cb)$ is strongly suppressed by the CKM matrix while the
decay $A\to b\overline b$ is normally the dominant decay for $A$. 

\subsubsection{Model I$'$ and II}
In Fig. 8 we plot the analogy of Fig.~7 for Model~I$'$.
We find a similar $\tan\beta$ dependence to that
 found for $H^{\pm}\to W^*h(A)$ in Fig. 4,
with a decrease in the BR as $\tan\beta$ increases. 
 The curve with $M_A=140$ GeV
allows the two--body decay $A\to hZ$, and maintains a large BR until 
$\tan\beta$ becomes very large. 

In Model~II
one may consider the decay $A\to hZ^*$, although since $M_{H^\pm}\ge 330$
 GeV in this model the $\rho$ parameter constraints (see Fig. 1)
suggest that requiring a light $h$ would cause $M_A$ to be out of the range
 of LEP2. For this reason we
do not plot a graph. Even if one allowed $M_A$ in range at LEP2 the
three--body decays would be very small, this being due to the fact that
the decay width for 
$A\to b\overline b$ is proportional to $\tan^2\beta$ in Model~II. 
If an on--shell $Z$ is allowed (e.g. $M_A=140$ GeV, $M_h=40$ GeV)
one finds BR($A\to hZ)\approx 70\%$
at $\tan\beta=2$, falling below $1\%$ at $\tan\beta=30$.
\begin{figure}[htb]
\centerline{\protect\hbox{\psfig{file=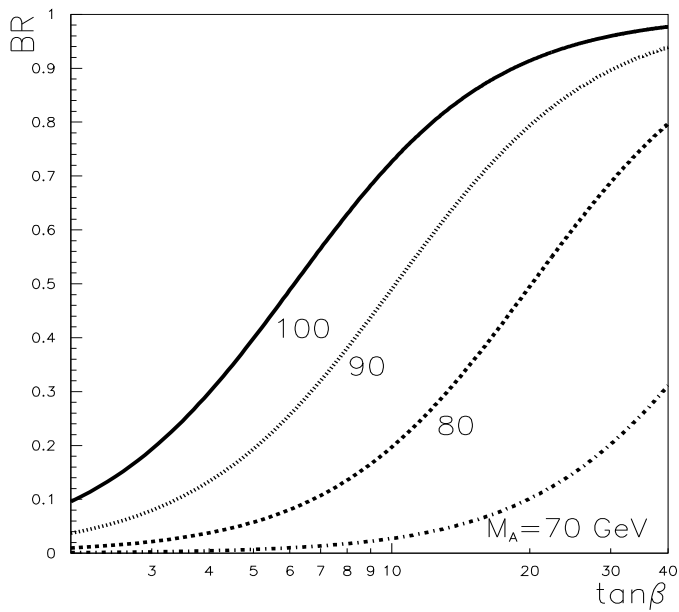,height=10cm,width=10cm}}}
\caption{BR$(A\to H^{\pm}W^*$) in Model~I against $\tan\beta$, for 
$M_{H}=55$ GeV.}
\label{cha6}
\end{figure}
\subsection{The decay $A\to H^{\pm}W^*$}
The decay $A\to H^{\pm}W^*$ may be considered in Models~I and I$'$.
 In Model~II the bound $M_{H^\pm}\ge 330$ GeV disallows this channel
 for $M_A$ in range at LEP2.
In Fig. 9 (for Model~I)
we vary $M_A$ from $70\to 100$ GeV, setting $M_{H^\pm}=55$ GeV. We
do not consider values of $M_A\ge 100$ GeV, since we require that the 
production process $e^+e^-\to hA$ be open, and also that
$M_h\ge M_A$ so that there are no competing $A\to hZ^*$ decays.
 We see that this channel
 can be dominant at larger values of $\tan\beta$.
For Model~I$'$ with $M_A=100$ GeV the BR peaks at $\approx 5\%$, and falls to
below $1\%$ for $\tan\beta\ge 10$ (we do not plot a graph).

\section{Detection in an enhanced $H^{\pm}cb$ channel}
We now consider the $H^{\pm}$ of the MHDM and aim to give a more 
detailed analysis of a detection channel mentioned in our earlier work. 
If no neutral Higgs boson exists with a mass lighter than
$M_{H^\pm}$ the BRs of $H^{\pm}$ in Model I and I$'$
will be given by the entries in Table 3.
In the MHDM there is a possibility of an enhanced $cb$ decay, which 
was first mentioned in Ref.~\cite{Gross} and in our earlier work \cite{Akestir}
we displayed the parameter space which allowed large
BR$(H^{\pm}\to cb)$.  We note that we did not use running masses of the
quarks in Ref.~\cite{Akestir}, and Figures 1 and 2 there are for 
$m_c=1.5$ GeV and $m_b=5$ GeV.
At the energy scale of 100 GeV these values drop to $m_c=0.62$ GeV and  
$m_b=3.04$
GeV, and so our previous results slightly overestimate BR$(H^{\pm}\to cb)$.
 However, it will still be possible to have a large BR$(H^{\pm}\to cb)$ 
$(\ge 10\%$) if $|X|\ge |Y|,|Z|$ by a factor of 5 or so.  

A review of the detection techniques for 
a light $H^{\pm}$ at LEP2 appears in Ref.~\cite{Sop}. The three signatures
analysed were $cscs$, $cs\tau\nu_{\tau}$ and $\tau\nu_{\tau}\tau\nu_{\tau}$. 
In the MHDM one could consider the signature $cb\tau\nu_{\tau}$,
which would give an isolated $\tau$ and missing energy recoiling
against a hadronic system with a tagged $b$ quark. 
Much of the selection
criteria would be identical to that of the $cs\tau\nu_{\tau}$ channel. 
In the $cs\tau\nu_{\tau}$ channel the $W^+W^-$ background is removed by
reconstructing the invariant masses of the $cs$ and $\tau\nu_{\tau}$ systems,
thus significantly reducing this background if $M_{H^\pm}\le M_W$.
The expected event numbers/efficiencies for the $cs\tau\nu_{\tau}$
channel are given in Table 5 (from Ref. \cite{Sop}). 
\begin{table}[htb]
\centering
\begin{tabular} {|l|c|} \hline
 Process & Eff. or No. of backg. evts  \\ \hline
 $H^{\pm}$ (60 GeV) & $5.6\%$  \\ \hline
 $q\overline q$& 0 \\ \hline 
 $W^+W^-$ & 2 \\ \hline
 $ZZ$ & 0 \\ \hline
$\tau^+\tau^-$ & 0 \\ \hline 
\end{tabular}
\caption{Expected signal efficiency and background event numbers in the
$cs\tau\nu_{\tau}$ channel for $\protect\sqrt s=175$ GeV.}
\end{table}

Detection of a $H^{\pm}$ will be very difficult in the 
region $M_{H}\approx M_W$, partly due to the lack of $H^+H^-$ pair production 
events for this mass region, and partly due to the fact that invariant mass
cuts which reduce the $WW$ background also remove the signal. In 
Ref.~\cite{Akestir} we
showed that the presence of a large BR$(H^{\pm}\to cb)$ (of order $50\%$
 is possible) in
a MHDM has two potential uses:
\begin{itemize}
\item [{(i)}] It would suggest that a detected $H^{\pm}$ is from the MHDM, 
since the equivalent decay in Model~I and Model~I$'$  has
a BR less than $1\%$.

\item [{(ii)}] It may allow a chance detection in the difficult 
$M_{H}\approx M_W$ region since $W^{\pm}$ rarely decays to $cb$.
\end{itemize}

The $cb\tau\nu_{\tau}$ channel
has the advantage of almost negligible background from $WW$ and $ZZ$
events, as long as one can correctly identify the $b$ quark. 
In addition we will be 
able to use the selection criteria of the $cs\tau\nu_{\tau}$ channel,
replacing the invariant mass cut with a $b$--tag requirement.
The number of signal events (before cuts) in the $cb\tau\nu_{\tau}$ 
channel is given by:
\begin{equation}
N_{H^+H^-}\times f(\rm BR)\,,  
\end{equation}
where $N_{H^+H^-}$ is the number of pair produced $H^{\pm}$ events, 
and $f(\rm BR)$ is defined by 
\begin{equation}
f(\rm BR)=2\times{\rm BR}(H^{\pm}\to cb)\times {\rm BR}(H^{\pm}\to
 \tau\nu_{\tau}) \,.
\end{equation}
\noindent
In order to isolate these final states $b$--tagging will be necessary.
Since this a standard technique for searching for the SM Higgs at LEP2,
the efficiency, $e_b$, will be quite high ($\approx 70\%$) in practice,
 see for example
Ref.~\cite{Aug}. For the $cb\tau\nu_{\tau}$ channel we shall use the
optimistic values BR$(H^{\pm}\to cb)={\rm BR}(H^{\pm}\to 
\tau\nu_{\tau})=50\%$
which maximizes $f(\rm BR)$. We note that in the $cs\tau\nu_{\tau}$ channel
 the invariant mass cut was the last cut applied, and all other
non--$WW$ backgrounds had already been removed. When it is applied for the
mass region $M_{H^\pm}\approx M_W$ it will remove the signal as well. Our aim
is to replace this latter cut with a $b$--tag requirement, which will 
reduce the
$WW$ background to negligible proportions while preserving most of the
signal. One would need to have a strong rejection against fake tags 
coming from $c$ quarks.
 From the figures in  Ref.~\cite{Sop} one can infer that 
  the invariant mass cut reduces the Higgs signal by $\approx 2/3$.
Therefore we shall assume a selection efficiency ($e_{H}$) before
$b$--tagging of $8.4\%$ for the
Higgs signal, obtained by scaling the value of $5.6\%$ in Table 5.
One can obtain the following formula for the number of signal events 
$(N_{sig})$
in the $cb\tau\nu_{\tau}$ channel:
\begin{equation}
N_{H^+H^-}\times e_H\times e_b\times f({\rm BR})\,.
\end{equation} 
The number of $cb\tau\nu_{\tau}$ events from $WW$ production is
1.1 {\sl before} any cuts have been applied. The cuts before
the $b$--tag requirement have a selection efficiency considerably
below $100\%$ and therefore the background is entirely
negligible. We stress that we require strong $c$ quark rejection since the 
number of 
events in the $cs\tau\nu_{\tau}$ channel is large. We then require 3 or more
 signal events for detection,
and we see from the Eq. (12) that the number of events for $M_{H^\pm}=80$ GeV
 and 
$\sqrt s=180$ GeV is equal to 2.1. Therefore detection is certainly marginal.
At the higher collider energy, $\sqrt s=200$~GeV, 
due to larger $N_{H^+H^-}$ we find that $N_{sig}=3.8$.
All this analysis is with optimistic choices for $e_b$ and $f({\rm BR})$. With 
greater luminosity, which would be available at a next generation collider
one could probe a greater parameter space of $f({\rm BR})$.

In summary, the $cb\tau\nu_{\tau}$ channel at LEP2 only provides a 
slight chance 
of overcoming the difficult $M_W\approx M_{H^\pm}$ region in the MHDM,
since the largest values of $f({\rm BR})$ would be needed.
However, the signature would have a use
for $M_{H^\pm}$ lighter than $M_W$ since it would provide evidence of 
the MHDM.
For $M_{H^\pm}$ comfortably below $M_W$ a reasonable number of $H^{\pm}$ pairs
would be produced and we shall
require three tagged $cb\tau\nu_{\tau}$ events to conclude that 
a detected $H^{\pm}$ originates from the MHDM. From Eq. (12)  we can
 obtain Eq. (13) from which the values of $f({\rm BR})$ needed to  
produce the distinctive signature of three tagged $cb\tau\nu_{\tau}$ events
can be found.
\begin{equation}
 f({\rm BR})\times N_{H^+H^-}=36\,. 
\end{equation}
Thus for $N_{H^+H^-}=100$ (corresponding to $M_{H}=75$ GeV at $\sqrt s=180$
GeV), one finds $f({\rm BR})=0.36$, and so BR$(H^{\pm}\to cb)\ge 20\%$
is required. For lower masses BR$(H^{\pm}\to cb)\approx 10\%$ (or even less)
would be sufficient. We note that for the 2HDM
 that the analogous signal would be $\le 0.1$ events, which
is unobservable. 

Another way of distinguishing $H^{\pm}$ of the
MHDM would be through a lack of $H^{\pm}\to \tau\nu_{\tau}$ decays
i.e. leptophobia with BR$(H^{\pm}\to jets\to 100\%$). 
From Table 1 one sees that BR($H^{\pm}\to \tau\nu_{\tau}$) is expected to
be large in both Model~I and I$'$ ($\approx 65\%$ and $\ge 95\%$
 respectively), while in the MHDM it can be reduced to much lower values. 
For example, a simple calculation shows (not including $cb$ decays)
\begin{equation}
 BR(H^{\pm}\to \tau\nu_{\tau}) \approx {1.8|Z|^2\over 1.8|Z|^2+|Y|^2} 
\end{equation} 
Therefore if $|Y|\ge 2|Z|$, one finds BR($H^{\pm}\to \tau\nu_{\tau})\le 30\%$.
Including the $cb$ decays would reduce this further, and so it is apparent
that a sizeable parameter space exists for BR($H^{\pm}\to \tau\nu_{\tau})\ll
65\%$. 
For the extreme case of BR$(H^{\pm}\to jets)\to 100\%$
one would find a ninefold increase in the number of events in the
$c\overline s\overline c s$ channel compared to
Model~I. 
 
\section{Conclusions}
We have studied the impact of three--body decays of a Higgs
boson to a lighter Higgs boson and a virtual vector boson in the context
of the non--supersymmetric 2HDM model.
Such decays have been studied in the MSSM, although their
importance is magnified in the four versions of the general 2HDM, partly 
due to the lack of correlation
 among both the mixing angles and masses of the Higgs scalars, and partly
due to the different couplings of the Higgs bosons to the fermions. 
Such decays would allow the production of light neutral Higgs bosons
which have so far escaped detection at LEP due to suppressed couplings
to the $Z$ boson. We showed that
 the three--body channels, if open, can be of great importance in
Model~I, permitting the decays
 $H^{\pm}\to h(A)W^*$ and $A\to hZ^*,H^{\pm}W^*$  
to proceed with large branching ratios over a wide range of $\tan\beta$ values.
 As $\tan\beta$ increases $H^{\pm}$ and $A$ decouple from the fermions and
 so the three--body channels rapidly grow in importance, 
 ultimately reaching
 branching ratios close to $100\%$. These results have important applications
 for the phenomenology of Model~I at LEP2, particularly for $H^{\pm}$ 
 which may avoid the present search techniques which assume charged
 scalar decays to $cs$ or $\tau\nu_{\tau}$. We suggested that the
 three--body decays might also allow detection of $H^{\pm}$ in the
 difficult $M_{H^{\pm}}$ region. 
 A fermiophobic Higgs boson  (only possible in Model~I) which has so far
 escaped direct searches in the $e^+e^-\to  H_FZ$ channel
  may be produced copiously in the decay of 
 $H^{\pm}$ or $A$, provided that enough of the latter are produced on--shell. 
 Signatures with 3 or 4 photons would be possible,
 which would pass
 current $H_F$ search criteria with equal or better efficiencies, and 
 enable discrimination from the $e^+e^-\to  H_FZ$ signal. For the 
 CP--even $h$ the three--body decays $h\to H^{\pm}W^*$
 and $h\to AZ^*$ may be dominant at low values of $\cos\alpha$ and
 allow a 6 fermion signature even when the decay $h\to AA$ is not open.
 
In Model~I$'$ the three--body decays can be significant (although not usually
dominant) for small values of $\tan\beta$. As $\tan\beta$ increases
the decays of $H^{\pm}$, $A$ and $h$ to the third generation of leptons
dominate. In Model~II one finds smaller branching ratios than the analogous
cases for Model~I$'$, and the decays involving $H^{\pm}$ are not relevant
at LEP2 energies due to the bound of $M_{H^{\pm}}\ge 330$ GeV.    
Finally, we showed that a $H^{\pm}$ in a general MHDM with $N$ doublets
may be distinguished from $H^{\pm}$ of a 2HDM if it possessed a sizeable
($\ge 10\%$) BR$(H^{\pm}\to cb)$ or a BR$(H^{\pm}\to jets)\gg 30\%$.

\section*{Acknowledgements}
This work was supported by DGICYT under grants PB95-1077, by the
TMR network grant ERBFMRXCT960090 of the European Union, and by a
CSIC--UK Royal Society fellowship. I wish to thank Marco Diaz 
and Andre Turcot for useful comments.


\begin{thebibliography}{99}
\bibitem{Arib} I wish to thank A. Arhrib for
pointing out that the formula displayed in Refs.~19 and 37
for $\Gamma(H^\pm\to W^*A(h))$ requires a extra multiplicative factor of 0.5.
\end{thebibliography}
\end{document}